\documentclass[10pt,twocolumn,a4paper,conference]{IEEEtran}

\usepackage{cite}	
\usepackage{graphicx} 
\usepackage[latin1]{inputenc} 
\usepackage[T1]{fontenc} 
\usepackage{amsmath,amsfonts,amsbsy,amssymb} 
\usepackage{mathabx} 
\usepackage{amssymb} 
\usepackage{amsmath}
\usepackage{amsthm}	
\usepackage{mathrsfs}
\usepackage[nolist]{acronym} 
\usepackage{tabularx} 
\usepackage{multirow}
\usepackage{wasysym}
\usepackage{float}
\usepackage{color} 

\usepackage{enumitem} 

\usepackage{graphicx}
\usepackage{multicol}

\newtheorem{remark}{Remark}

\begin{document}

\title{Fixed Rate Statistical QoS Provisioning for Markovian Sources in Machine Type Communication} 
\author{ Fahad Qasmi, Mohammad Shehab, Hirley Alves, and Matti Latva-aho\\
	
	\IEEEauthorblockA{
		Centre for Wireless Communications (CWC), University of Oulu, Finland\\
	}
	Email: firstname.lastname@oulu.fi
}
%
\maketitle


\begin{abstract}
In this paper, we study the trade-off between reliability and latency in machine type communication (MTC), which consists of single transmitter and receiver in the presence of Rayleigh fading channel. We assume that the transmitter does not know the channel conditions, therefore it would be transmitting information over a fixed rate. The fixed rate transmission is modeled as a two-state continuous-time Markov process, where the optimum transmission rate is obtained. Moreover, we conduct a performance analysis for different arrival traffic originated from MTC device via effective rate transmission. We consider that the arrival traffic is modeled as a Markovian process namely Discrete-Time Markov process, Fluid Markov process, and Markov Modulated Poisson process, under delay violation  constraints. Using effective bandwidth and effective capacity theories, we evaluate the trade-off between reliability-latency and identify QoS (Quality of Service) requirement, and derive lower and upper bounds for the effective capacity subject to channel memory decay rate limits.
     
\end{abstract}
	

\section{Introduction}\label{introduction}
\vspace{-.3mm}
The next generation of  communication systems will bring new business models and revolutionary changes in several industry chain processes due to ultra-reliable low latency communication (URLLC) and massive MTC (mMTC) use cases, enabling the industrial  Internet of Things (IoT). As a result, there are foreseeable  benefits in terms of efficiency, safety and sustainability for the industry and society. Therefore, it can be easily inferred that MTC is one of the core elements in the IoT revolution that will be required for achieving ambitious goals planned for future  wireless networks \cite{P1}.

Ultra-reliable low latency communication (URLLC) aims to  provide communication services  where stringent QoS requirements are imposed in terms of reliability   and latency. This type of communication is normally used in mission critical applications such as vehicle-to-vehicle communication (V2V), remote surgery and critical links between industrial processes.  On the other hand, mMTC refers to the kind of network services that can support large number of connected smart devices. It can be used to cover monitoring, automation and infrastructure of buildings, smart agriculture, logistics, tracking and fleet management. The devices are normally used to collect and forward information in both real as well as non-real time modes. These connected devices are expected to be greater than 50 billion by 2020\cite{2020}.
 
MTC (mMTC and URLLC) has attracted much interest in the recent years which promises a huge potential and market growth for the technology. Furthermore, research and development suggest that cellular networks are suitable to largely satisfy the requirements of MTC in terms of global reach, QoS, scalability, security and diversity. A traffic model is a stochastic process that matches the behaviour of physical quantities of measured data traffic. Current cellular networks are based on the standard traffic model, which is designed and optimized for typical behaviour of human subscribers. Moreover, the traffic behaviour of MTC is different,  since it is mostly uplink dominant; uses short  as well as less number of packets; is (Non) real-time, periodic and event driven. It is usually bursty in nature (suddenly the volume of data flow increases in response to trigger of certain events) and coordinate in nature (i.e. simultaneous access attempts from many machines reacting to the same events), while HTC (Human Type Communication) is uncoordinated one. The QoS requirements of MTC are different  in terms of security, reliability and latency\cite{mtraffic-issue}.

%

Consequently, there is a prominent need for traffic models that can not only capture the behaviour of MTC traffic but can also provide the adequate communication services for both types of traffic with required QoS. Traffic models are mainly characterized into source and aggregated traffic models. The source traffic model captures the behaviour of individual users or sources. These models are often based on Poisson process  which is an appropriate choice for capturing the properties of individual MTD (Machine Type Devices) generated traffic.
On the other hand, the aggregate traffic model where a large number of MTD are assigned to one aggregator for capturing the traffic pattern of an individual user due to it's homogenous and coordinated nature of the traffic \cite{Mtraffic12}.


The key design issue for traffic generated by MTD is how to achieve acceptable performance and  efficiently use delay bound guarantees  for timely transmission within the specified QoS requirements in a wireless channel. Wireless channel is highly time varying channel due to the random changes in the environment or obstacles. These changes in wireless signal usually lead to variations in the strength of the received signal, which  directly affects the QoS guarantee in a MTC. Therefore, deterministic delay bound QoS constraints are usually difficult to satisfy. Alternatively, the statistical QoS provisioning is a powerful tool to characterize and implement delay bound QoS guarantees for wireless real time traffic\cite{HQoS1}.

In this context, effective capacity is a statistical QoS provisioning metric and can be defined as the maximum constant arrival rate that a given time varying service process can support while providing statistical latency guarantees \cite{main}. It is derived from the large deviation theory and incorporates the statistical QoS constraint by capturing the decay rate of the buffer occupancy probability for the queue length. Effective capacity ensures the QoS guarantee on the constant arrival rates. Here, we are particularly interested to conduct a throughput analysis of random and bursty source traffic patterns by using Markovian source models including discrete time Markov, Markov fluid and Markov modulated Poisson sources with effective capacity.

Effective capacity \cite{Shehab2018} has extensively been used over the past few years to evaluate the trade-off among the reliability, latency, security and energy efficiency. For example, the authors  \cite{liela} evaluate the trade-off between reliability and energy efficiency under QoS constraints. In \cite{r7}, an optimal power allocation scheme is proposed to maximize the energy efficiency under given QoS constraint. In \cite{hirley_s1}, the authors assess the performance of security in MTC networks using  secure statistical QoS provisioning. In \cite{main5}, the authors have considered fixed-rate transmission modelled as a two-state (ON/OFF) continuous-time Markov chain and utilize effective capacity to analyse energy efficiency.


In this paper, we conduct performance analysis of optimum fixed rate model. By using effective capacity theory, we aim at meeting the adequate reliability and latency requirement in point-to-point MTC network. We follow the recent contribution in \cite{my1} to incorporate Markovian arrival traffic so that its impact on the optimum transmission rate performance of the network can be evaluated. We derive upper and lower bounds of the effective capacity for high and low channel memory values. Our work is different from \cite{my1} as we assume a fixed rate transmission over Rayleigh fading channel which is modelled as continuous time Markov chain. This model allows us to identify the level of reliability and latency that each transmission possesses and therefore design the appropriate optimum transmission rate by formulating an optimization problem, that not only maximizes the effective capacity but also increases the allowed maximum average arrival rate at sources. Hence the satisfactory reliability and latency requirement are achieved under statistical QoS constraints which are satisfied by the system.

\section{Preliminaries}
\subsection{System Model}
In this paper, we consider single transmitter and receiver in the presence of Rayleigh fading. The input output relation of the channel model can be expressed as \vspace{-1mm} 
\begin{align}\label{eq1}
y(t)=h(t)x(t)+n(t),
\end{align}
where $x(t)$ and $y(t)$ are the complex valued  input and output signals, respectively, and  $n(t)$ is zero mean, circularly-symmetric, complex Gaussian noise. Finally, $h(t)$ is  the Rayleigh  fading coefficient which denotes the multiplicative fading component expressing the attenuation and phase shift experienced in the channel, and it is assumed to be  a zero-mean complex Gaussian process. Therefore, $z(t)$ = $|h(t)|^2$ has an exponential distribution and assume unit variance.

In this paper, we assume that the receiver is able to estimate the channel coefficient $h(t)$, whereas the transmitter does not know this information. Therefore the transmitter would be transmitting information over a fixed rate $\cal R$ bits/s.
The wireless channel changes slowly and hence, $h(t)$ stays  constant over each coding block, the instantaneous channel capacity of Gaussian channel model is given by
\begin{align}\label{eq2}
C(t)=\rm log_2(1+\gamma \mathit{z(t)} )\  bits/s,
\end{align}

where $\gamma$ is the average transmitted signal-to-noise ratio. When $\cal R$ $<C(t)$, then the channel is considered to be in ON state. Hence, the transmitted message is decoded
correctly and reliable communication is accomplished. While $\cal R$ $\geq C(t) $, then the channel is considered to be in OFF state;  outage arises and retransmission is needed.


The  transition rate from ON to OFF state is denoted by $\mu$, and OFF to ON state is $\nu$ as illustrated in Fig. 1. Then, we can write that the channel  ON state probability is equivalent to  $ Pr \{z(t)\! >\! \Psi \}=\int_{\Psi}^{\infty}e^{-z}\ dz=e^{-\Psi}=\frac{\nu}{\nu+\mu}$, where $\Psi=\frac{2^{\cal R}-1}{\gamma}$, and the OFF state probability is as $Pr \{z(t)\! \leq\! \Psi\}=\int_{0}^{\Psi}e^{-z}\ dz =1-e^{-\Psi}=\frac{\mu}{\nu+\mu} $. Hence, we have $\nu=\kappa e ^{-\Psi}$ and $\mu=\kappa(1-e^{-\Psi})$. These equations are necessary to determine $\nu$ and $\mu$. It is noted that the channel memory of these two-state Markov process  exponentially decay at  rate $\nu+\mu=\kappa$\cite{main5}. 
\begin{figure}[!t] 
	\centering \vspace{-2mm}
	\includegraphics[width=0.7\columnwidth]{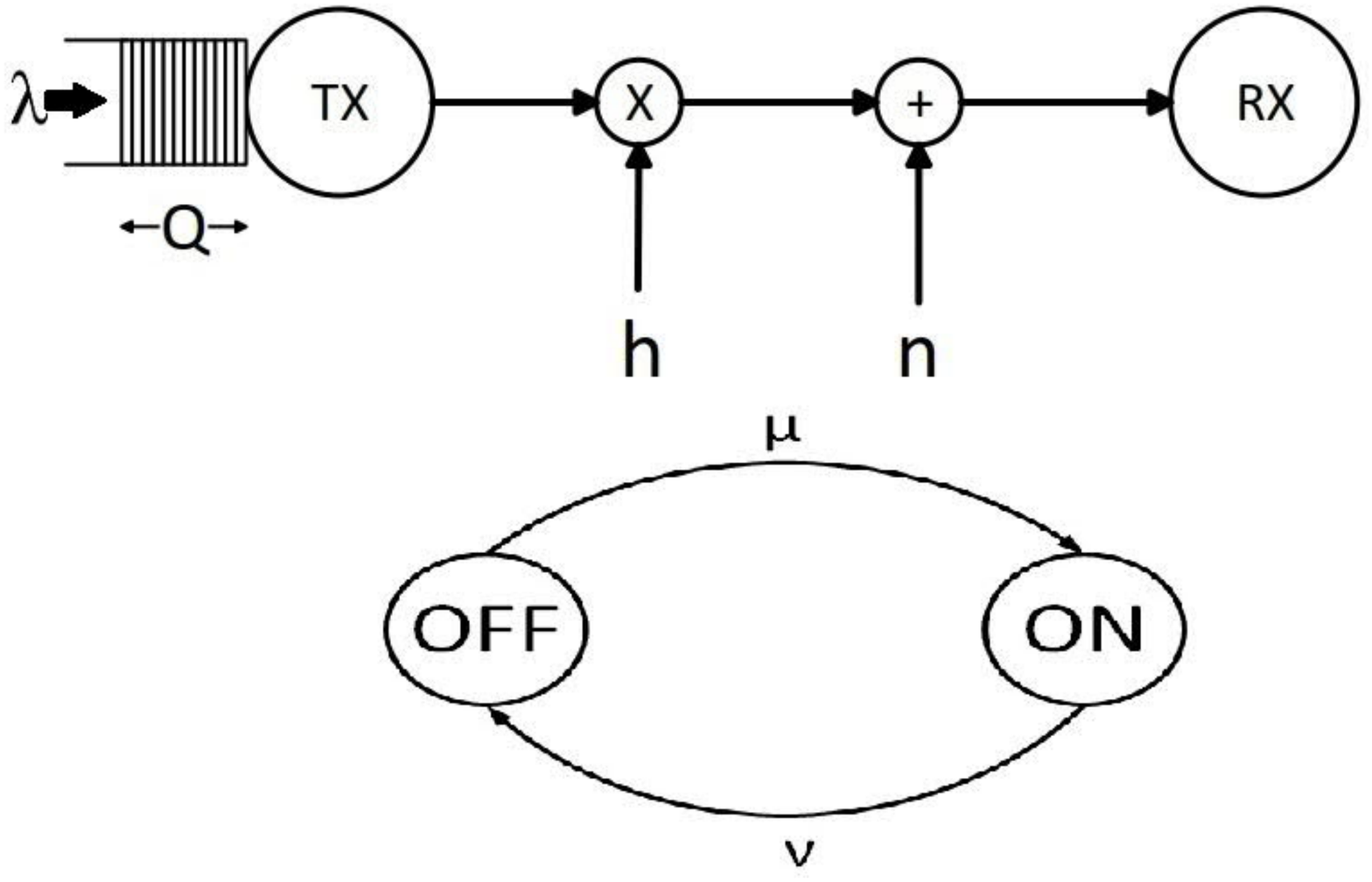}
	\centering \vspace{-0mm}
	\caption{System Model, where $\lambda$ is the arrival rate at the source, and $Q$ is the length of the queue at the transmitter. The channel coefficient is denoted as $h$ and AWGN noise as $n$. Transmission in wireless link is also modeled as a continuous-time Markov chain with ON and OFF states, whose transition rate of fading states  are measured  by $\nu$ and $\mu$.}
	\vspace{-3mm}
	\label{f2}
\end{figure}


\subsection{Throughput of delay constrained networks}
It is assummed  that before transmission, data generated by random sources is stored  in  First in First out (FIFO)  buffer  as illustrated in Fig. \ref{f2}. Thus,  delay may occur in transmitting the data  because of the long waiting time of data in the buffer. Moreover, the delay overflow probability is given by \cite{main}
\begin{align}\label{eq3}
\text{Pr}\{D\geq d\}\approx\zeta e^{-\theta a(\theta) d},
\end{align}
where $D$ is the queueing delay, $d$ is the delay threshold, $\theta$ is the delay QoS constraint, $a(\theta)$ is the effective bandwidth and $\zeta$ is the probability of non-empty buffer. Larger value of $\theta\rightarrow \infty$ implies that stringent QoS constraint is imposed, while for loose QoS constraint the value $\theta\rightarrow 0$ is small \cite{shehab}.

In this paper, we  consider three types of Markov arrival sources namely Discrete-Time Markov source (DTMS), Fluid Markov source (FMS) and Markov Modulated Poisson source (MMPS). These Markov sources are using  two-state ON and OFF model. In  ON state, data arrives   with rate $\lambda$ bits/block and the OFF state refers to no arrival of data as illustrated in Fig. \ref{f2}. For these sources, effective bandwidth provides a mean to characterize the minimum constant service rates required to support the random arrival of data into the buffer constrained to some statistical QoS requirements, namely buffer violation probability in \eqref{eq3}. Let the time accumulated arrival process at instant $t$ be $A(t)=\sum_{k=1}^{t} a(k)$. Then the effective bandwidth is defined as \cite{main}
\begin{align}\label{eq4}
a(\theta)=\lim_{t\rightarrow\infty}{1\over \theta t}\log\mathbb{E}\{e^{\theta A(t)}\} \ ,
\end{align}

Effective capacity ($C_E$) is the dual concept of effective bandwidth, where it defines the maximum constant arrival rate that a given time-varying service process can support in order to guarantee a statistical QoS requirement specified by the QoS exponent $\theta$. The effective capacity for a given QoS exponent is obtained from \cite{120}
\begin{align} \label{eq17}
C_{E}({ \gamma}, \theta)=-\lim_{t\rightarrow\infty}{1\over \theta t}\log\mathbb{E}\{e^{-\theta S[t]}\},
\end{align}
where $S[t]\triangleq \sum_{k=1}^{t} P[k]$ is the time accumulated service process and $\left\{P[k],k=1,2,\cdots\right\}$ shows the discrete time stationary and ergodic stochastic service process. So  $P[k]=\cal R$, when it is ON state and otherwise 0 in OFF state. 

We formulate a two-state continuous-time Markov chain fixed rate effective capacity model for a given statistical constraint, where  $\theta$ as
\begin{align} \label{17}
C_{E}({\gamma},\theta,{\cal R})={1\over 2\theta}\left[\theta{\cal R}+(\nu+\mu)- \xi\right],
\end{align}
where $\xi=\sqrt{(\theta{\cal R}+(\nu+\mu))^{2}-4\nu\theta {\cal R}}$, and \eqref{17} represents a simplified version of the effective capacity.

\begin{remark}
	\label{lm:misome_pso}
	The upper bound of effective capacity is expressed as below
	\begin{align}
	C_{E}=\lim_{\kappa \rightarrow \infty}{1\over 2\theta}\left[\theta{\cal R}+(\nu+\mu)- \xi\right]={\cal R}e ^{-\Psi},
	\end{align}	
	describes that $C_E$ is asymptotic at the high channel memory decay rate  towards ${\cal R}e ^{-\Psi}$, which is free from QoS constraints $\theta$. Meanwhile, there is a lower bound for $C_E$, which approaches zero for very low channel memory decay rate as follows
	\begin{align}
	C_{E}=\lim_{\kappa \rightarrow 0}{1\over 2\theta}\left[\theta{\cal R}+(\nu+\mu)- \xi\right]=0.
	\end{align}
\end{remark}

It is fairly hard to investigate the performance of the communication system when the arrival of data and channel characteristics are random in nature with certain target guarantees.  We assume the maximum average arrival rate is the throughput metric of Markovian arrival sources that can support wireless fading channel and satisfy QoS guarantees illustrated in (3). Thus, QoS constraints are fulfilled when the effective bandwidth of the arrival process is equal to the effective capacity of service \cite{main}, therefore
\begin{align}\label {eq11}
a(\theta)=C_{E}( \gamma,\mathrm{\theta, \cal R}),
\end{align}
Then we can find maximum arrival rate that can support fixed rate transmissions at given  $\gamma$ and $\theta$. 

\section{Maximum arrival rate of Markovian Sources}
\subsection{Discrete - Time Markov Sources}
In the DTMS model, the arrival of data in the buffer is discrete in time. We assume a simple two-state ON/OFF model. In ON state $\lambda$ bits arrive in the buffer, whereas no arrival of data in OFF state. Effective bandwidth in this case is defined by \cite{my1} 

\begin{align}\label{eq5}
a(\theta)&={1\over \theta}\log_{e}\left({1\over 2}\left(p_{11}+p_{22}e^{\theta  \lambda}\right. \right. \notag \\
&\left.\left.+\sqrt{(p_{11}+p_{22}e^{\theta \lambda})^{2}-4(p_{11}+p_{22}-1)e^{\theta \lambda}}\right)\right), 
\end{align}
where $p_{11}$ denotes the probability of staying in OFF state, while $p_{22}$ determines the probability of ON state. The transition probabilities from one state to another are denoted by $ p_{21}=1-p_{22} $ and $ p_{12}=1-p_{11}$.
$\mathrm{P_{ON}}$ is the probability of ON state in the steady state regime, which is used to the calculate average arrival rate as
\begin{align}\label {eq6}
\lambda_{\mathrm{avg}}= \lambda\cdot\mathrm{P_{ON}}=\lambda \ \frac{1-p_{11}}{2-p_{11}-p_{22}}\ ,
\end{align}
when the queue is in steady state then arrival rate is equal to the departure rate  \cite{my1}.

We substitute the effective bandwidth expression of discrete time Markov source \eqref{eq5} in \eqref{eq11} and solve as follows
\begin{align}\label{eq12}
(\rm p_{11}+p_{22}{e}^{\lambda \theta }-2{e}^{\theta C_{E}( \gamma,\theta)})^2&=(\rm p_{11}+p_{22}{e}^{\lambda\theta })^2 \\ \notag
&-4(\rm p_{11}+p_{22}-1){e}^{\lambda\theta }.
\end{align}
After solving \eqref{eq12} for $\lambda$, we obtain maximum ON state arrival rate as
\begin{align}\label{eq13}
\lambda^{*}(\theta)\!=\!\frac{1}{\theta}\log_{e}\!\left(\!\frac{{e}^{2\theta C_{E}( \gamma,\theta)}-p_{11}{e}^{\theta C_{E}( \gamma,\theta)}}{(1-p_{11}-p_{22})\!+\!p_{22}{e}^{\theta C_{E}( \gamma,\theta)}}\!\right).
\end{align}
By incorporating \eqref{eq6} we can express the maximum average arrival rate as a function of QoS exponent, effective capacity fading channel and state transition probabilities as \cite{my1}
\begin{align}\label{eq14}
\lambda_{\mathrm{avg}}^*( \gamma,\theta)\!=\!\frac{\mathrm P_{\text{ON}}}{\theta}\log_{e}\!\left(\!\frac{{e}^{2\theta C_{E}( \gamma,\theta)}-p_{11}{e}^{\theta C_{E}( \gamma,\theta)}}{1\!-p_{11}\!-p_{22}+p_{22}{e}^{\theta C_{E}(\gamma,\theta)}}\!\right)\!.\!
\end{align}

Let us further simplify our analysis by assuming $p_{11}=1-s$ and $p_{22}=s$, we obtained a simplified version of the source model in which single parameter $s$ measure the burstiness of the source \cite{main}.

\begin{align}\label{2_eq261}
\lambda_{\mathrm{avg}}^*( \gamma,\theta)=\frac{ s}{\theta}\log\left(\frac{ {e}^{\theta C_{E}(\theta)}-(1-s)   } {   s }\right).
\end{align}

\subsection{Markov Fluid Sources}
In this model, the data continuously arrive in the buffer and the effective bandwidth is expressed by \cite{main}  
\begin{align}\label{eq7}
a(\theta)\!=\!{1\over 2\theta}\left[\theta \lambda\!-\!(\!\alpha+\!\beta)\!+\!\sqrt{(\theta \lambda-(\alpha+\beta))^{2}+4\alpha\theta \lambda}\right]\!,
\end{align}
where $\alpha$ shows the transition rate from OFF state to ON state and $\beta$ is the transition rate from ON state to OFF state. Then, we attain the steady state probability of being ON as
\begin{align}\label {eq8}
\mathrm{P_{ON}}=\frac{\alpha}{\alpha + \beta }.
\end{align}
We follow a similar procedure as for the DTMS to determine the maximum average arrival rate of two-state ON/OFF model for the MFS as \cite{my1}
\begin{align}\label{eq15}
\lambda_{\mathrm{avg}}^*(\gamma,\theta)=\mathrm{P}_{\text{ON}}\frac{\theta C_{E}(\gamma,\theta)+\alpha+\beta}{\theta C_{E}(\gamma,\theta)+\alpha}C_{E}(\gamma,\theta).
\end{align}
\subsection{Markov Modulated Poisson Sources}
In Markov modulated Poisson process, the arrival of data in the buffer is a Poisson process, whose intensity is controlled by a continuous time Markov chain. In OFF state, there is no arrival of data which means arrival intensity is zero. On the other hand, $\lambda$ is the arrival intensity in ON state. In this case the effective bandwidth is defined as \cite{main}

\vspace{-5mm}
\begin{align}\label{eq10}
\begin{split}
a(\theta)&={1\over 2\theta}\left[\ (e^{\theta}-1) \lambda-(\alpha+\beta)\right]\\
&+{1\over 2\theta}\sqrt{(\ (e^{\theta}-1) \lambda-(\alpha+\beta))^{2}+4\alpha (e^{\theta}-1) \lambda}.
\end{split}
\end{align}
Similar to the previous sources models, we determine the maximum average arrival rate for the MMPS source model as \cite{my1} \vspace{-2mm}
\begin{align}\label{eq16}
\!\lambda_{\mathrm{avg}}^*(\gamma,\theta)\!=\! \mathrm P_{\text{ON}}\frac{\theta[\theta C_{E}(\gamma,\theta)  +  \alpha+ \!\beta]}{(e^\theta-1)\theta C_{E}( \gamma,\theta) +\alpha}C_{E}(\rm \gamma,\theta).
\end{align}

\section{Optimal Fixed Rate transmission}
\begin{figure}[!h] 
	
	\centering
	\includegraphics[width=1\columnwidth]{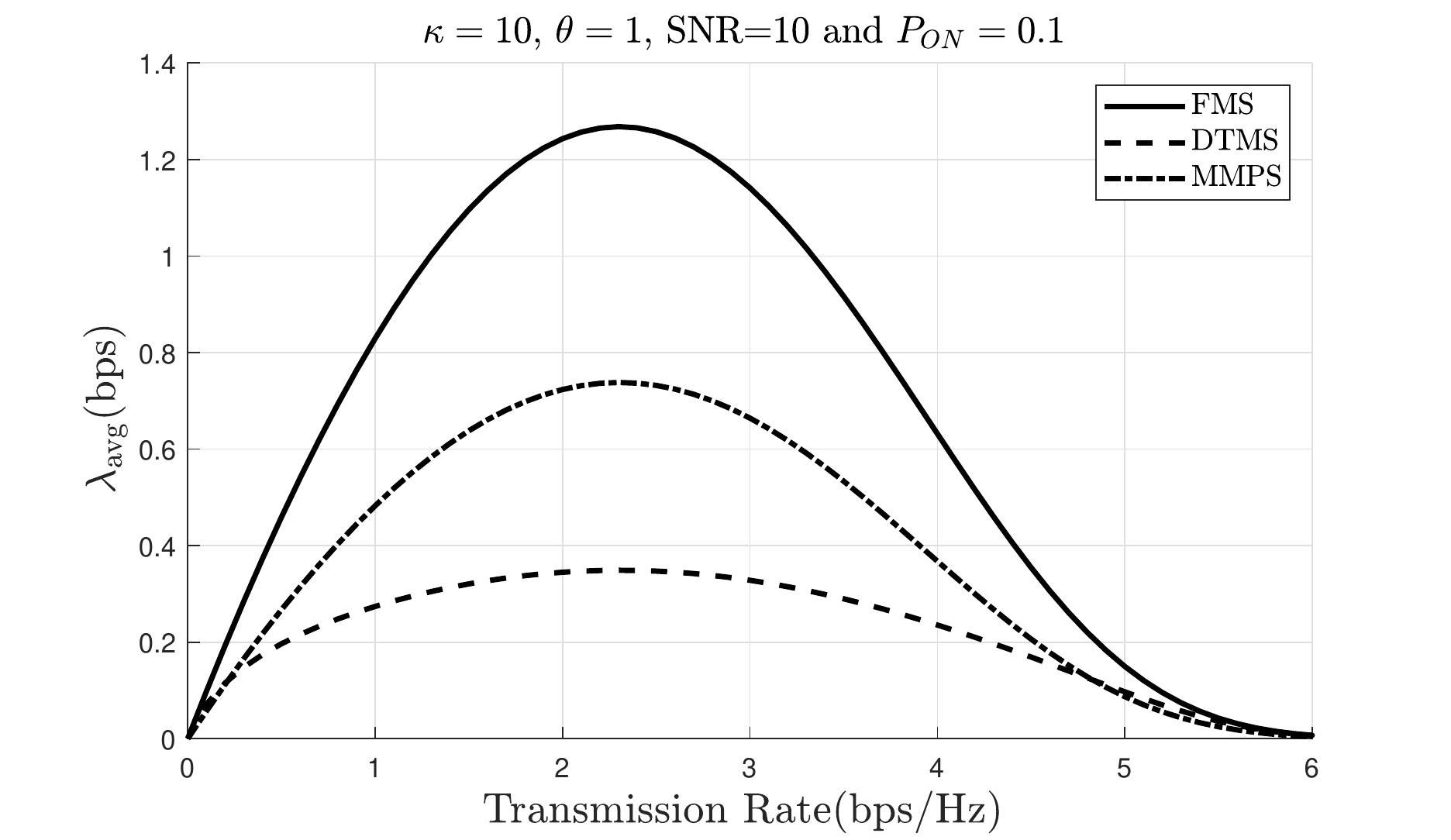}
		\caption{Maximum average arrival rate of Markovian sources as a function of transmission rate ${\cal R}$ for fixed values  of QoS constraint  $\theta $,  $\kappa$, $\gamma$ and $\mathrm{P_{ON}}$.}
	\label{f22} \vspace{-2mm}
\end{figure}

Fig. $\ref{f22}$ shows the maximum average arrival rate of Markovian sources as a function of transmission rate ${\cal R}$ for fixed values  of QoS constraint  $\theta $,  $\kappa$, $\gamma$ and $\mathrm{P_{ON}}$. It is clearly visible that the maximum average arrival rate is maximized  as transmission rate increases, but after certain limit the effective capacity gradually starts to decrease due to fixed value of $\gamma$. If one transmits with higher rate, it will degrade the overall throughput. Therefore, efficient use of transmission rate boosts the performance of communication system. Furthermore, it is noticed that optimum transmission rate which maximizes the $\lambda_{\mathrm{avg}}$ has similar value for all markovian models. We conclude that optimizing the effective capacity with respect to the transmission rate, allows for high link throughput while allowing larger arrival rates as we shall see next.

The optimization problem to maximize $C_E$ subject to a non-negative transmission rate
\begin{align} \label{20}
C^*_{E}({\gamma},\theta,{\cal R} )=\max_{{\cal R}\geq 0} \left\{ {1\over 2\theta}\left[\theta {\cal R}+(\nu+\mu)-\xi\right]\right\}
\end{align}
Now, taking first derivative of \eqref{20} and equating   it to zero yields the optimum value of ${\cal R}$.
\begin{align} \label{21}
\frac{\partial C^*_{E}({\rm \gamma}, \theta, {\cal R})  }{\partial {\cal R}}=\frac{\theta-\frac{2 \theta({\cal R}\theta+\kappa)-4\theta \nu + \frac{2^{2+{\cal R}} {\cal R} \log(2) \nu} {\gamma}} \xi  }{2\theta}=0.
\end{align} 
\noindent After some algebraic manipulation, we have
\begin{align} \label{27}
\gamma \ \xi -\gamma ({\cal R}\theta+\kappa)+2\gamma \nu-2{\cal R}\nu\log(2)\ 2^{\cal R}=0
\end{align}
Further, we simplify \eqref{27} to reach
\begin{align} \label{23}
{\cal R^*}=\frac{\gamma(\xi - \kappa+2 \nu)}{\gamma \theta+2\nu \log(2)2^{\cal R^*}}.
\end{align}
However, closed form solution for (24) does not exist. In order to obtain optimum transmission rate ${\cal R^*}\!$, we use fzero Matlab root function.
\section{results and discussion}

\begin{figure}[h] 
	\vspace{-2mm}
	\centering
	\includegraphics[width=1\columnwidth]{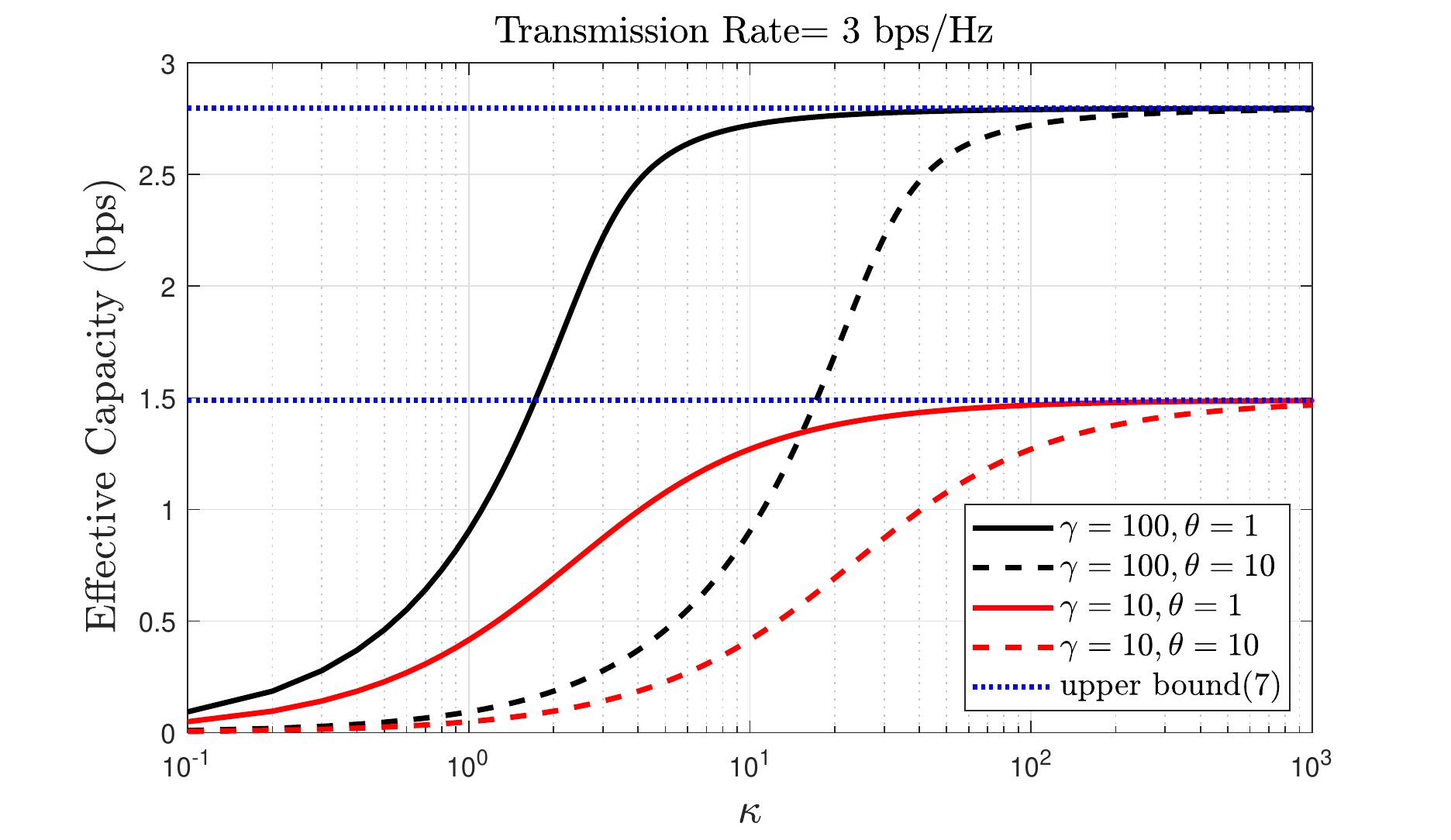}
	\caption{Effective capacity as a function of $\kappa$ for different values of $\gamma$ $(\mathrm{SNR})$ and $\theta$ when $\cal R$$=3$.}
	\label{f11} \vspace{-2mm}
\end{figure}

Herein, Fig. $\ref{f11}$ shows the effective capacity as a function of channel memory $\kappa$ for different value of $\gamma$ and $\theta$, when $\cal R$$=3$ bps. High values of  $\kappa$ measure  the channel memory decays fast which corresponds to good channel state and bears higher transmission rate but has less probability of this event occurred. We observe that the effective capacity is maximized at high level of channel memory rate for  low $\gamma$ curve as compared to high $\gamma$ when stringent QoS constraint is imposed and then it saturates. It is shown in Remark 1 that channels with high channel memory $\kappa$ produce an asymptotic boost in effective capacity up towards an upper bound, which is  physical phenomena that can measure with $\kappa$ but do not control. We conclude that, larger channel memory $\kappa$ or $\gamma$ above a certain level will not increase  effective capacity.

\begin{figure}[h] 
	\vspace{-0mm}
	\centering
	\includegraphics[width=1\columnwidth]{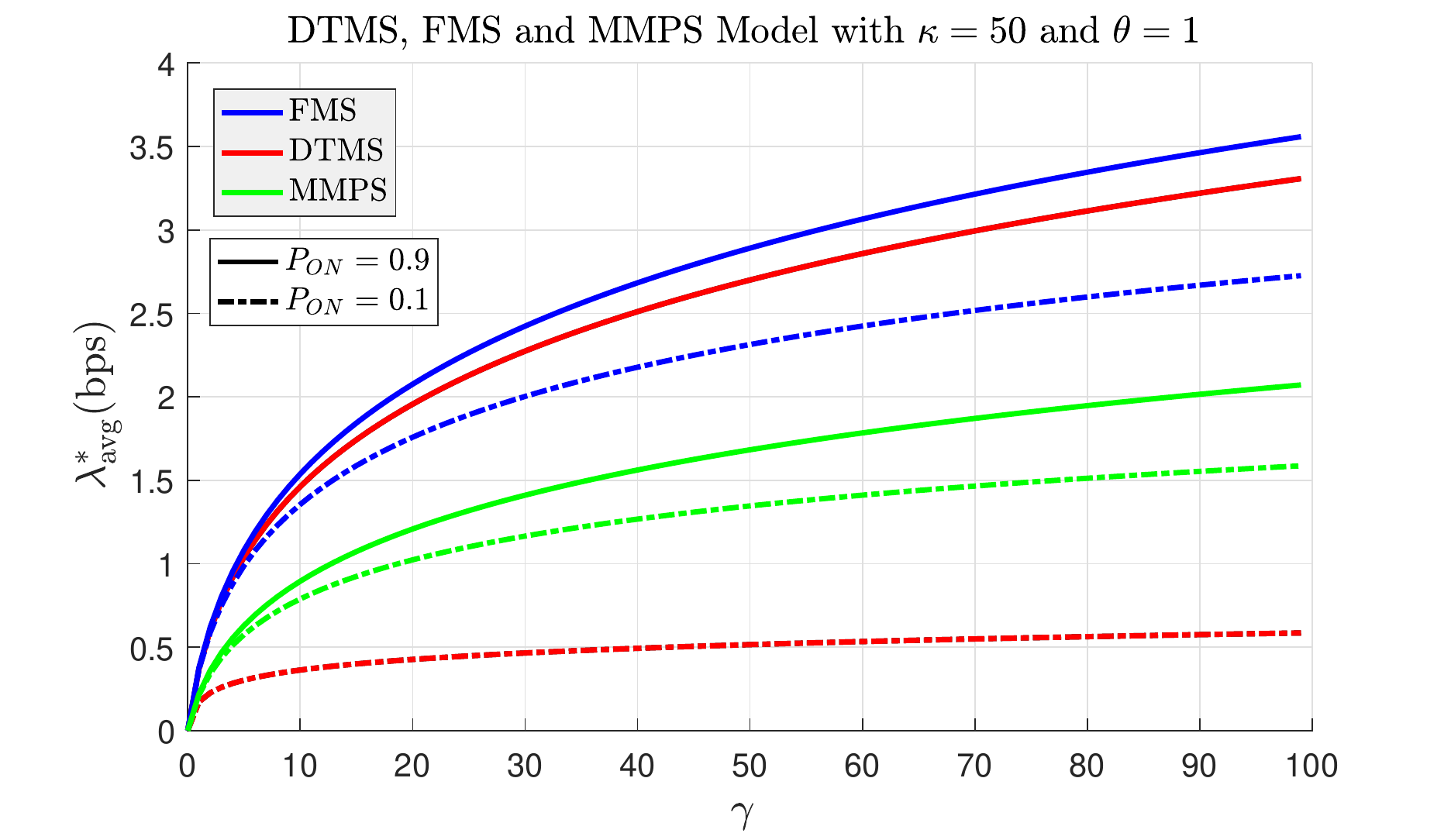}
	\caption{Maximum average arrival rate $\lambda^*_{\mathrm{avg}}$ as a function of $\gamma$ $(\mathrm{SNR})$  for distinct Markov source models with different value of $\mathrm{P_{ON}}$.}
	\label{f33}
	\vspace{-3mm}
\end{figure}

In a wireless communication system, it is generally presumed that by increasing signal power, the performance of the system  improves. Fig. $\ref{f33}$ illustrates the maximum average arrival rate $\lambda^*_{\mathrm{avg}}$ of Markovian models as a function of $\gamma$ for different values of $\mathrm{P_{ON}}$ when we set QoS exponent $\theta=1$ and $\kappa=50$. It is clear that as the $\gamma$ increase throughput also improves. The increase in burstiness of sources (i.e, when $\mathrm{P_{ON}}$ decreases) implies that the data arrives less frequently which means that the source is bursty. Moreover, the arrival rate in ON  state becomes higher.  

%

\begin{figure}[h] 
	\vspace{-1mm}
	\centering
	\includegraphics[width=1.02\columnwidth]{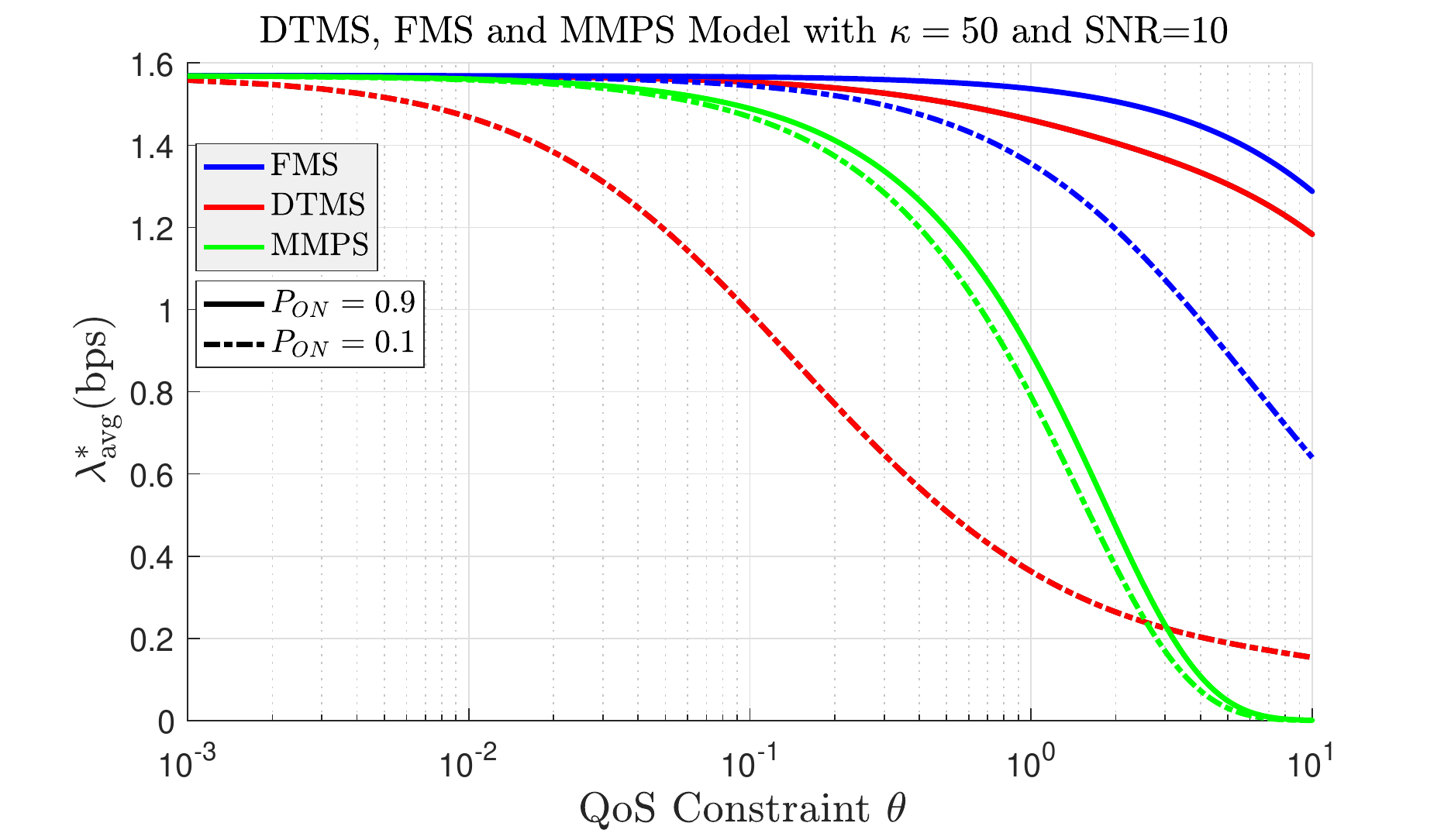}
	\caption{Maximum average arrival rate $\lambda^*_{\mathrm{avg}}$ as a function of $\theta$  for distinct Markov source models with different value of $\mathrm{P_{ON}}$.}
	\label{f44}
\end{figure}

The low value of $\theta$ is examined in  Fig. $\ref{f44}$, we plot the maximum average arrival rate $\lambda^*_{\mathrm{avg}}$ as function of QoS constraint $\theta$ for distinct Markov source models with different value of source characteristic $\mathrm{P_{ON}}$ when we set $\gamma=10$ and $\kappa=50$. It is noted that higher value of $\lambda_{\mathrm{avg}}$ can be sustained for the loose QoS constraints. Hence, throughput does not depend  on channel memory and source characteristics. When the QoS constraints becomes too stringent, the throughput tends to zero, since assigned arrival rate  must satisfy short delay requirement. This  reduction in $\lambda^*_{\mathrm{avg}}$ is more severe in bursty sources (i.e, when $\mathrm{P_{ON}}=0.1$). Moreover,  maximum average arrival curve of MMPS sources is generally smaller and declines fast, when stringent QoS constraint are imposed. This can be the natural attributed to the much more randomness/burstiness tolerated at high $\gamma$ regime for the DTMS and FMS. Furthermore, MMPS throughput is less affected by burstiness because MMPS sources exhibit a higher level of variation in this sense and can be regarded as a more bursty source.

\begin{figure}[h] 
	\centering
	\includegraphics[width=1.02\columnwidth]{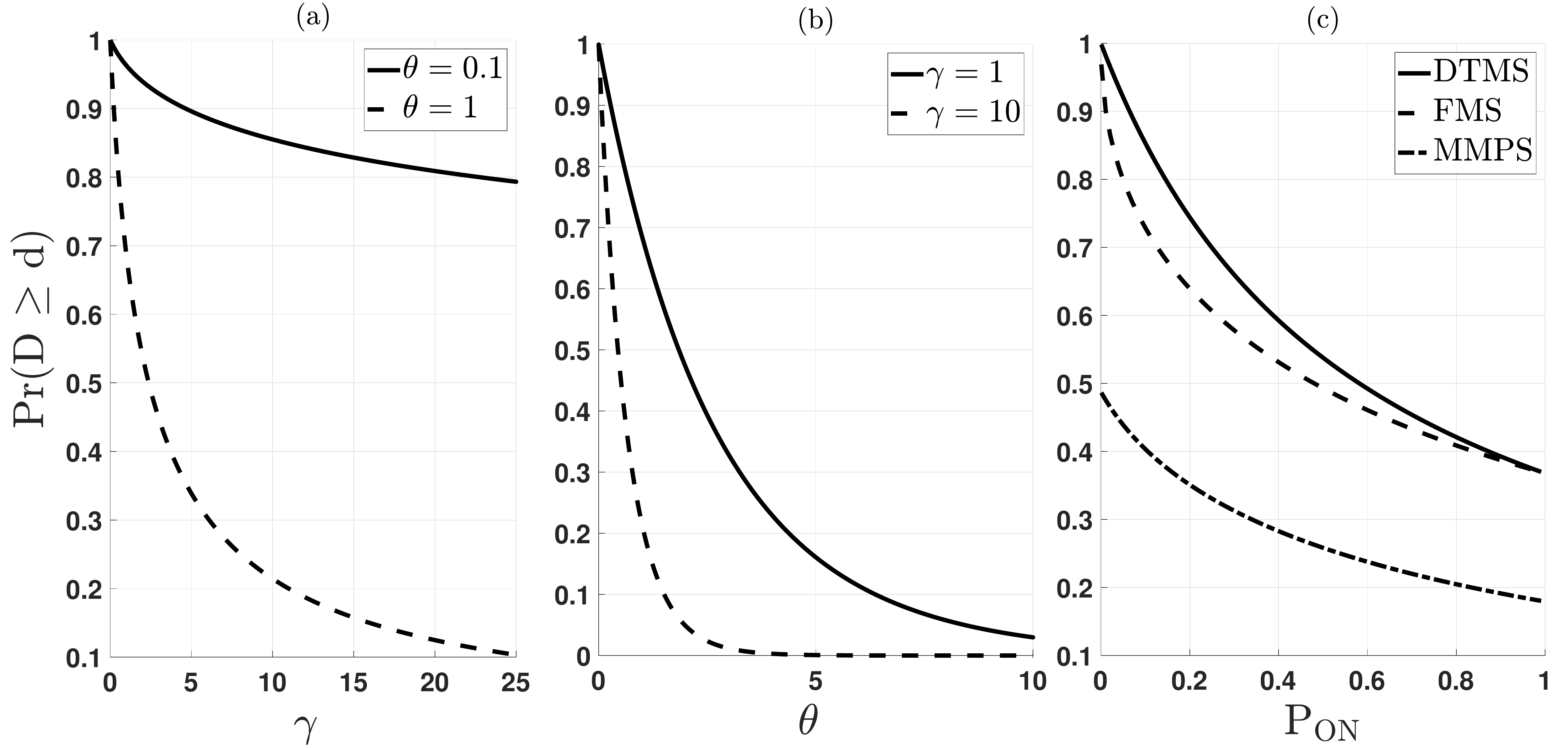}
	\caption{Delay violation probability as function of  (a) $\gamma$ assuming fixed $\theta$, (b) $\theta$ assuming fixed $\gamma (\mathrm{SNR})$ and (c) $\mathrm {P_{ON}}$.}
	\label{f55}
	\vspace{-1mm}
\end{figure}
Fig. $\ref{f55}$ includes subplots for the outage that occurs due to delay violation probability as a function of $\gamma$, QoS exponent $\theta$ and source characteristic $\mathrm{P_{ON}}$. It is obvious that the reliability and latency tradeoff is described in terms of delay violation probability. Enhancing reliability requires large values of $\gamma$, tight QoS requirements and sources should behave less bursty. The relation between reducing latency and enhancing reliability is a trade-off subject to the design requirements. In Fig. \ref{f55}(a), as $\gamma$ increases, good channel quality is available for optimal transmitting rate, which  tends to  boost reliability. Fig. \ref{f55}(b) shows that as QoS requirement increases, the waiting time of data in buffer shorten which uplifts reliability level. In Fig. \ref{f55}(c), delay violation probability is plotted as a function of $\mathrm {P_{ON}}$ for fixed arrival rate of 1 bps and fixed QoS requirements $\theta=0.01$. Burstiness is measured from the average arrival of data in ON state. It is observed that more bursty sources degrade the maximum average arrival rate, which directly rises delay violation probability. Furthermore, the   MMPS sources tolerate lower delay violation probability.  Hence, the reliability of MMPS is greater as compared to DTMS and FMS in the presence of random/bursty sources.

\begin{figure}[!t] 
	\vspace{-2mm}
	\centering
	\includegraphics[width=1\columnwidth]{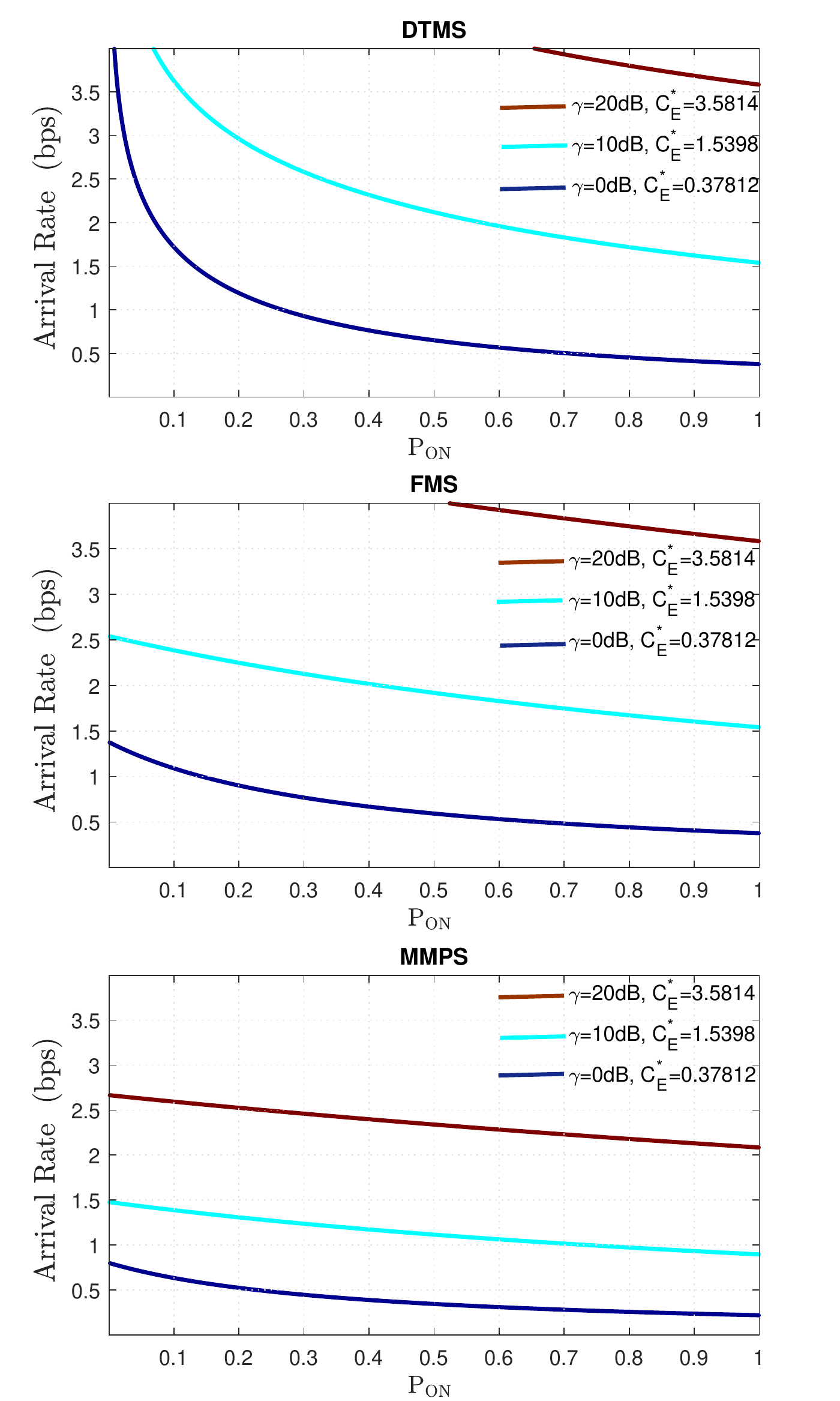}
	\caption{Arrival rate vs $\mathrm{P_{ON}}$ for different $\mathrm{C_E^*}$ of Markovian sources.}
	\label{f66}
	\vspace{-3mm}
\end{figure}

In Fig. $\ref{f66}$, we investigate the impact of optimum effective capacity $C_E^*$ on the random/bursty sources under QoS Constraints. It is clearly observed that bursty sources (i.e., less $\mathrm {P_{ON}}$ ) have higher arrival rate in ON state because of maintaining  average arrival rate nondecreasing function as defined  \eqref{eq6}. However, with the same departure rate, it is challenging to keep throughput non-decreasing when QoS constraints are imposed. Therefore, the higher arrival rate is required to achieve $C_E^*$ when the source becomes bursty (i.e.$\mathrm{P_{ON}<1}$). It is also observed that due to more bursty sources MMPS has a small rise in arrival rate required to support given $C_E^*$ as compared to DTMS and FMS. Thus MMPS is more stable and convenient for modelling bursty or random source traffic.
\vspace{-2mm}
\section{Conclusions} \label{conclusions}
\vspace{-0.5mm}
In this work, we conducted the detailed performance analysis of effective transmission rate model to achieve adequate reliability and latency trade-off in single point-to-point machine type devices. Effective transmission rate model is based on effective capacity for continuous time markov chain model. We characterized the upper and lower bounds of effective capacity as a function of channel memory decay rate and defined the optimum fixed transmission rate which maximizes the throughput for Markovian source models. The results showed that source, buffer, channel memory and transition rates have major impact on the system performance when certain reliability and latency constraints are imposed. Furthermore, increased source burstiness and stringent QoS requirement all need an increase in SNR and decay rate of channel memory to fulfill the requirement of reliable communication. Moreover, it is observed that MMPS model is more stable for more bursty sources when compared to FMS and DTMS in the presence of QoS constraints. 
\section*{Acknowledgments}
This work is partially supported by Academy of Finland 6Genesis Flagship (Grant no. 318927), Aka Project EE-IoT (Grant no. 319008), and by Finnish Funding Agency for Technology and Innovation (Tekes), Bittium Wireless, Keysight Technologies Finland, Kyynel, MediaTek Wireless, and Nokia Solutions and Networks.

 \bibliographystyle{IEEEtran}
 \bibliography{fahad2}

\end{document}